   \documentstyle[12pt,twoside,epsf]{article}

\begin{document}
%
%
\begin{titlepage}
  \noindent{\large\tt DESY 95-022 \hfill ISSN 0418-9833}
  \vspace{-.45cm}
\begin{flushleft}
  ITP-UH-06/95 \\ hep-ph/9502324  \end{flushleft}
  \vspace{-1.4cm}
\begin{flushright}  February 1995 \quad \end{flushright}
\begin{center}  \vfil \vspace{1.6cm}
{\Large \bf
  The Infrared Sensitivity of \\[12pt]
  Screening and Damping in a Quark-Gluon Plasma }
\vfil  \vspace{1.2cm}
{\large
 F. Flechsig$^a$ , \ A. K. Rebhan$^b$\footnote{On leave of
    absence from Institut f\"ur Theoretische Physik der
    Technischen Universit\"at Wien} \ and \ H. Schulz$^a$ } \\
\end{center}  \medskip \smallskip \qquad \qquad
{\sl $^a$} \parbox[t]{12cm}{ \sl
  Institut f\"ur Theoretische Physik, Universit\"at Hannover, \\
  Appelstr. 2, D-30167 Hannover, Germany\\ } \\
\bigskip \qquad \qquad
{\sl $^b$} {\sl DESY, Theory Group, Notkestr.
                85, D-22603 Hamburg, Germany}
\vspace{1cm} \vfil
%
%
\centerline{\large  ABSTRACT}
  \begin{quotation}
All the next-to-leading order contributions to the quasi-particle
dispersion laws of a quark-gluon plasma which due to infrared
singularities are sensitive to the magnetic-mass scale are
calculated using Braaten-Pisarski resummation. These
relative-order-$g\ln (g)$ corrections are shown here to generally
contribute to the dynamical screening of gluonic fields with
frequencies below the plasma frequency as well as to the damping
of propagating gluonic and fermionic quasi-particles. In the limit
of vanishing wave-vector the infrared singularities disappear, but
in a way that raises the possibility for formally higher orders
of the Braaten-Pisarski scheme to equally contribute at
next-to-leading order when the wave-vector is of the order of or
less than the magnetic-mass scale. This is argued to be a problem
in particular for the fermionic damping rate.
   \end{quotation} \vspace{.4cm}
\end{titlepage}

%
%

\vspace{1.5cm} \hspace{.08cm} \parbox{15cm}{{\bf
1. \ Introduction }
\vspace{1.1cm} } \hfill \vphantom{a} \nopagebreak \indent

The leading-order results for the dispersion laws of
quasi-particles in a quark-gluon plasma \cite{KKW} are well
known and are readily obtained from the high-temperature limit
of one-loop Green's functions or from solving the collisionless
Boltzmann equation \cite{Silin}. However, anything beyond the
leading terms becomes accessible only through a resummation of
the conventional perturbation series.

A systematically improved perturbation theory, which turns
out to involve single powers of the QCD coupling constant $g$
rather than the usual $g^2$, has been established some years
ago by Braaten and Pisarski \cite{BP}. It is based on a
resummation of all the leading-order self-energies and
vertex functions furnished by the so-called hard thermal
loops \cite{BP,FT}. In a first application this led to the
solution of the long-standing problem \cite{samm} of how to
calculate the damping constant of the lowest QCD plasmon
excitation with vanishing wave vector in a gauge-invariant
way \cite{BPpd}. In later work it was found that
with non-zero wave vector the damping of gluonic as well as
fermionic excitations is infrared divergent with a
logarithmic singularity which can be cut off by a finite
screening mass for static magnetic fields. This gives rise to
a contribution proportional to $\ln(m_{\rm el.}/m_{\rm magn.})
\sim \ln(1/g)$ whose coefficient is calculable perturbatively
\cite{BMAR,LS2,APG,HP,Pmov}. Similar singularities have also
been found recently in the next-to-leading order calculation of
the nonabelian Debye screening mass \cite{AKR,AKR2} from the
pole of the static gluon propagator (at imaginary wave vectors)
as well as in the perturbative evaluation of the correlator of
Polyakov loops \cite{AKR2,BrN}.

Such a sensitivity to the magnetic mass scale comes as a
surprise since by superficial infrared power counting
\cite{Linde} one would not expect it already at (resummed)
one-loop order. It is in fact due to the necessity to evaluate
the loop diagrams at the position of the pole of the
leading-order propagator, which leads to ``mass-shell''
singularities in the presence of the massless magnetostatic
mode. Indeed, these singularities appear also in the case
of QED, where no magnetic screening mass can be generated.
Presumably the finite width $\gamma$ of the full propagators
is also able to provide the required infrared cutoff \cite{LS2},
which would again lead to $\ln(m_{\rm el.}/\gamma) \sim
\ln(1/g)$, or $\ln(1/e)$ in the case of QED.

In this work we shall present a unified treatment of all these
next-to-leading order corrections to gluonic and fermionic
dispersion laws which are infrared singular due to mass-shell
singularities (Section 2) and therefore sensitive to the
`magnetic-mass scale', by which we mean simply the scale of
new physics that acts as an effective infrared cutoff, be it
an actual magnetic mass or the shielding of the singularities
through a finite width of the full propagators or
another mechanism. For moving quasi-particle excitations
we reproduce the results of Pisarski \cite{Pmov} who found
that the infrared-singular contributions are proportional to
the group velocity of the respective modes. For the case of
dynamical screening of perturbations with frequencies below
the plasma frequency we find direct proportionality to the
modulus of the (now imaginary) wave vector (Sections 3 and 4).
We also take a somewhat closer look (Section 5) at the limit of
vanishing wave-vector where the infrared singularities seem to
be absent and we shall argue that for wave-vectors of the order
of or smaller than the magnetic-mass scale these infrared
singularities could still leave non-negligible imprints which
would render the resummation scheme of Braaten and Pisarski
incomplete in this limit.

%
%

\vspace{1.5cm} \hspace{.08cm} \parbox{15cm}{{\bf
2. \ Quasi-particle mass-shell singularities }
\vspace{1.1cm} } \hfill \vphantom{a} \nopagebreak \indent

The leading-order finite-temperature corrections to the gluon
and fermion self-energies \cite{KKW} give rise to effective
(albeit momentum-dependent) thermal masses of the order $gT$,
where $g$ is the QCD coupling constant and $T$ the temperature.
Because of a nontrivial tensor structure, they give rise to
different dispersion laws for spatially longitudinal and
transverse gluons according to
\begin{eqnarray} \label{ldl}
  Q^2 &=& {\mit\Pi}_\ell(Q_0,q) \nonumber \\
  &=& - \; {Q^2 \over q^2} \,{\mit\Pi}_{00}
     \; =\; 3m^2 \left( 1 - {Q_0^2 \over q^2} \right)
     \left( 1 - {Q_0 \over 2q} \ln\left( {Q_0 + q
     \over Q_0 - q} \right) \right) \\
 \label{tdl}
  Q^2 &=& {\mit\Pi}_t (Q_0,q) \nonumber \\
  &=& {1\over 2} \left( 3m^2-{\mit\Pi}_\ell \right)
\end{eqnarray}
where ${\mit\Pi}_{\mu\nu}$ is the gluon self-energy and
$m^2 = (N+N_f/2) (gT/3)^2$ for gauge group SU($N$) with $N_f$
flavors. For fermions (whose zero-temperature rest-mass we
assume to be $\ll gT$), there are also two different modes at
finite temperature from the solution of
\begin{equation} \label{det}
  \det \left( Q\!\llap{/} - \Sigma \right) = 0 \;\; ,
\end{equation}
where $\Sigma$ is the fermion self-energy. Eq.~(\ref{det}) has
two solutions for positive $q$, which are given by
\begin{eqnarray} \label{fermdl}
   Q_0 &=& \pm q + {\mit\Pi}_\pm(Q_0,q) \\
  \hbox{with}&& {\mit\Pi}_\pm = {M^2\over 2q}
  \,\left[\, \left( 1 \mp {Q_0\over q} \right)
  \ln \left( {Q_0 + q\over Q_0 - q}\right) \pm 2 \,\right]\,
  \end{eqnarray}
and $M^2=C_F(gT)^2/8$, $C_F=(N^2-1)/(2N)$.

In Fig.~1, the dispersion laws $Q_0=\omega(q)$ of the above
modes are shown as curves in the $\omega^2,q^2$-plane. For real
$q$, $q^2>0$, they correspond to propagating quasi-particles;
for $q^2<0$, they describe the screening of fields oscillating
below the plasma frequency, where $|q|$ is the inverse
screening length.

The physical significance of the various modes have been
discussed in full length by Weldon \cite{KKW}, to which we
refer for further detail. Let us just mention here that the
spatially longitudinal gluonic excitation (which is often
referred to as {\em the} plasmon) and the fermionic one
corresponding to the lower sign in Eq.~(\ref{fermdl})
(sometimes called the plasmino) have no counterpart at zero
temperature. They can only be understood as collective
phenomena.

 \begin{figure}                                            
     \centerline{ \epsfxsize=6in                           
     \epsfbox[85 310 500 485]{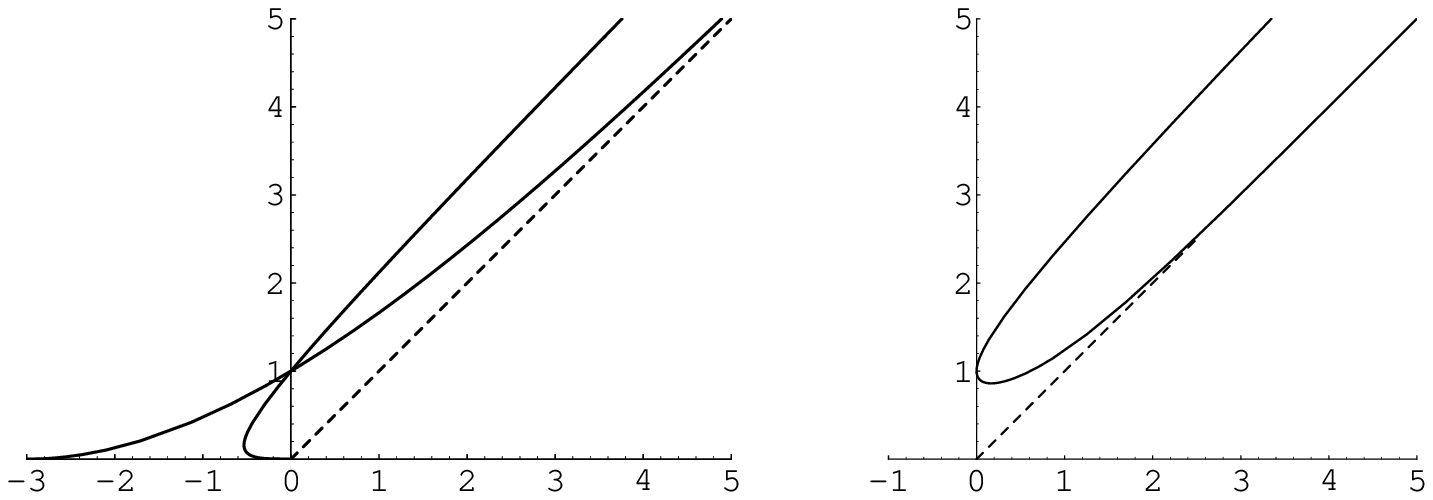} }                    
 \vspace{-6cm}  \font\frm=cmr10
 \unitlength1cm \begin{picture}(16,5)
 \put(3.5,5.0){$\omega^2$} \put(10.75,5.0){$\omega^2$}
 \put(1.7,4.1){{\frm IR-singular}}
 \put(1.7,3.6){{\frm screening}}
 \put(4,4.1){{\frm IR-singular}}
 \put(4,3.6){{\frm damping}}
 \put(11.25,4.1){{\frm IR-singular}}
 \put(11.25,3.6){{\frm damping}}
 \put(11.72,2.6){$_+$}    \put(12.5,2){$_-$}
 \put(4.8,2.5){$t$}       \put(5.15,2.2){$\ell$}
 \put(2.1,.4){$\ell$}     \put(8.6,-.15){$q^2$}
 \put(15.85,-.15){$q^2$}  \end{picture} \\[30pt]
 \parbox[t]{1cm}{$\phantom{a}$} \parbox[t]{1.7cm}{{\frm
  Figure 1$\,$:$\;$}}  \parbox[t]{12.2cm}{{\frm
The leading-order gluonic and fermionic dispersion curves and
their infrared singularities at subleading order. For the
transverse ($t$) and longitudinal ($\ell$) dispersion curves
of the gluonic modes $\omega^2$ and $q^2$ are given in units
of $m^2$; for the quark ($+$) and plasmino ($-$) modes the
unit is $M^2$.}}
\end{figure}                                               

If one attempts to go beyond the leading-order results just
presented, one has to face the problem that the conventional,
bare perturbation theory becomes insufficient whenever the loop
momenta probe the new (``soft'') scale $\sim m, M$. This is
true in particular for the next-to-leading order terms in
Eqs.~(\ref{ldl},\ref{tdl},\ref{fermdl}). An improved perturbation
theory, due to Braaten and Pisarski \cite{BP}, requires a
resummation of both the leading-order contributions to the
various self-energies and also to the vertices, collectively
termed `hard thermal loops'. However, in a perturbative treatment
of the solutions to Eqs.~(\ref{ldl},\ref{tdl},\ref{fermdl})
one has to evaluate the correction terms to the self-energies at
the location of the poles of the leading-order propagators, which
in the presence of massless modes generally will give rise to
``mass-shell'' singularities. Indeed, while all the propagating
physical modes have nonvanishing effective masses, the thermal
mass of the transverse gluon propagator vanishes in the static
limit, corresponding to the absence of screening of static
chromomagnetic fields up to distances $\sim 1/(gT)$.

Consider the following integral which appears in the resummed
one-loop contribution to the self-energy of mode $i$, where in
the loop integral the propagator for mode $i$ is coupled to
the one of transverse gluons,
\begin{eqnarray} \label{S}
  {\cal S}_i(Q_0,q) &=&
  \sum_{P_0=2\pi inT}\int{d^3p\over (2\pi)^3}\,\Delta_t(P)\,
  \Delta_i(Q-P)\,\Big|_{\Delta_i^{-1}(Q)=0} \nonumber \\
  &\equiv & \sum_{P_0=2\pi inT}\int{d^3p\over (2\pi)^3} \;
  {1\over P^2 - {\mit\Pi}_t(P)}\; {1\over (Q-P)^2 -
  {\mit\Pi}_i(Q-P)} \;\Bigg|_{Q^2 = {\mit\Pi}_i(Q)} \;\; .
\end{eqnarray}
For gluons, $i$ either means $t$ or $\ell$; for fermions
the second propagator in (\ref{S}) has the form
\begin{equation}
  {1\over \,\left[\, K_0 - k - {\mit\Pi}_+ (K) \,\right]\,
  \,\left[\, K_0 + k - {\mit\Pi}_- (K)\,\right]\, }
  \;\,\equiv\;\, \Delta_f (K)
\end{equation}
with $K\equiv Q-P$.

The $n=0$ contribution in the above sum comes with a vanishing
${\mit\Pi}_t(P_0=0,p)=0$, and for small $p$ the second
propagator is $\sim 1/({\bf pq})$ on the mass-shell of mode
$i$, so this contribution involves a logarithmic singularity
on mass-shell. Concentrating on the infrared-singular part,
we therefore have
\begin{eqnarray} \label{st}
  {\cal S}_i (Q_0,q) &\simeq& T\int{d^3p \over (2\pi)^3} \;
  {1\over p^2} \; {1\over p^2 - 2{\bf pq}
  + {\mit\Pi}_i(Q-P) - {\mit\Pi}_i(Q) - i\epsilon \sigma\,}
  \;\Bigg|_{P_0=0} \hspace{3cm} \nonumber \\
  &\simeq& T \,\left[\, 1 + \partial_{q^2} {\mit\Pi}_i(Q_0,q)
    \,\right]^{-1} \int{d^3p\over (2\pi)^3}\; {1\over p^2}\;
    {1\over p^2 - 2{\bf pq} -i\epsilon \sigma'\,} \;\; , \quad
\end{eqnarray}
where here and in the following $\simeq$ indicates that we have
dropped regular terms. In the last line of Eq.~(\ref{st}) we
have used that the difference of the two ${\mit\Pi}_i$ when
taken at $P_0=0$ is a function of $Q$ and $p^2-2{\bf pq}$. Only
the first term of its Taylor series with respect to the latter
variable is relevant for the infrared singular contribution;
the others lead to regular integrals of the form
$\int\! d^3p \, p^{-2} (p^2 -2 {\bf pq})^n$, $n\ge 0\,$.
In Eq.~(\ref{st}) $\sigma$ is a sign fixed in accordance to
retarded boundary conditions by $Q_0\to\omega + i\epsilon$:
\begin{equation}
  \sigma = {\rm sign} \left( \omega
  \,\left[\, 1 - \partial_{\omega^2}{\mit\Pi}_i(\omega,q)
  \,\right]\, \right)
\end{equation}
which leads to
\begin{equation}
  \sigma' = {\rm sign} \left( \omega \,\left[\,
  1 - \partial_{\omega^2}{\mit\Pi}_i(\omega,q) \,\right]\, /
  \,\left[\, 1 + \partial_{q^2} {\mit\Pi}_i(\omega,q)
  \,\right]\, \right) = {\rm sign}\left(v_i\right)
\end{equation}
with $v_i$ the group velocity $d\omega/dq$ of mode $i$. Note
that $-\omega^2 \partial_{\omega^2} = q^2 \partial_{q^2}$
since ${\mit\Pi}_i(\omega,q)$ depends only on $\omega/q$.

In order to give meaning to the singular expression (\ref{st}),
we introduce an infrared cut-off $\lambda\ll gT$ for
$p\,$:
\begin{equation}
  {\cal S}_i ( \omega , q\, ; \lambda )
  = {T\over 4\pi^2} \,\left[\, 1 + \partial_{q^2}
    {\mit\Pi}_i(\omega,q) \,\right]^{-1}
    \int_\lambda^\infty{dp\over p}\; {1\over 2q}
    \,\ln\left( p + 2q - i\epsilon \sigma' \over  p - 2q
    - i\epsilon \sigma' \right) \;\; ,
\end{equation}
which has a regular real part but a singular imaginary part,
where $\lambda$ cannot be removed,
\begin{equation} \label{srq}
   {\cal S}_i(\omega,q\, ;\lambda)
   = i\sigma' {T\over 8\pi q} \,\left[\, 1 + \partial_{q^2}
     {\mit\Pi}_i \,\right]^{-1} \ln\left( {q \over \lambda}
     \right) + O \left( \lambda^0 \right) \;\; .
\end{equation}

For frequencies below the plasma frequency $m$, where the
gluonic dispersion laws describe screening corresponding to
poles at purely imaginary values of $q=\pm i|q|$, we have
instead
\begin{eqnarray} \label{siq}
  {\cal S}_i(\omega,q=\pm i|q|;\lambda)
  &=& {T\over 4\pi^2} \,\left[\, 1 + \partial_{q^2} {\mit\Pi}_i
   \,\right]^{-1} \int_\lambda^\infty{dp\over p}\;
  {1\over |q|} \,\arctan\left( 2|q|\over p\right) \nonumber\\
  &=& {T\over 8\pi |q|} \,\left[\, 1+\partial_{q^2} {\mit\Pi}_i
   \,\right]^{-1} \ln\left( {|q|\over \lambda}\right)
  + O \left( \lambda^0 \right)
\end{eqnarray}
which now has a singular {\em real} part.

The results (\ref{srq},\ref{siq}) are only valid when
$|q|\gg\lambda$. In the limit $|q|\to0$, $\cal S$ is in fact
linearly divergent rather than only logarithmically, which
makes $\cal S$ proportional to $1/\lambda$:
\begin{equation} \label{slin}
  {\cal S}_i(\omega,q;\lambda) \to {T\over 4\pi^2\lambda}
  \,\left[\, 1 + \partial_{q^2} {\mit\Pi}_i \,\right]^{-1}
  + O \left( {Tq\over \lambda^2}\right) \quad
  \hbox{when $|q|\ll\lambda$}  \;\; .
\end{equation}

In the applications below it will turn out that all the infrared
singular contributions to the next-to-leading order
self-energies involve rational functions of the momenta
times the paradigmatic expression $\cal S$ such that the
limit $q\to0$ is infrared finite and, apparently, the
linear divergence in (\ref{slin}) is defused. However, we shall
find reason to reassess the case $q\to0$ more cautiously later,
so let us for the time being
restrict our attention to the case of a non-zero $q$.

With non-zero $q\sim gT$, the infrared singularities are only
logarithmic and it suffices to assume that an actual infrared
cutoff will be generated at some scale $\mu\sim g^x T$, $x>1$ to
make sense of the calculations. The technical cut-off $\lambda$
can then be identified with $\mu$ as far as the logarithmic term
is concerned, which turns into $(x-1)\ln(1/g)$.
We expect $\mu\sim g^2T$ and shall refer to it loosely as the
magnetic-mass scale, but actually we mean simply the mass scale
where new (infrared) physics eventually removes the mass-shell
singularities of Eq.~(\ref{st}). This may be through the
nonperturbative generation of a screening mass for static
chromomagnetic fields or something else, e.g. the finite width
of the true quasi-particle excitations \cite{LS2}, which, as
we shall see, is again of the order of $g^2T$ (up to
logarithms $\ln(g)$).

At any rate, this assumption fixes the term involving $\ln(1/g)$,
while the finer details of the physics pertaining to the
supersoft scales, which are outside the scope of this paper,
will be relevant only for the subleading contributions.

%
%

\vspace{1.5cm} \hspace{.08cm} \parbox{15cm}{{\bf
3. \ Gluon self-energy }
\vspace{1.1cm} } \hfill \vphantom{a} \nopagebreak \indent

In resummed perturbation theory, the one-loop correction to the
gluon self-energy is usually a rather complicated object, since
also the vertices have to be dressed. For non-zero external
wave-vector the resulting algebra is considerable. Some of its
properties have recently been studied by two of the present
authors \cite{flesh}. Fortunately, when concentrating on the
infrared-singular parts, this algebra can be greatly simplified
and we shall in the following describe only the shortcut to
obtaining them. In an appendix we display the next-to-leading
order self-energy in full detail, from which the following
results can be straightforwardly reproduced, albeit in a less
transparent manner.

First notice that only the diagrams containing two propagators
are able to produce mass-shell singularities, of which one has
vanishing 4-momentum. The most singular terms are those that
have no further loop momenta in the numerators, so all the
momentum algebra factorises into a prefactor to singular
integrals of the form (\ref{S}). In particular, only the part
of the complicated dressed vertices contributes where one leg
has vanishing momentum. Thanks to the gauge invariance of
the hard-thermal-loop effective action \cite{BP}, these vertices
are determined by differential Ward identities,
\begin{equation}
  {\,}^*\!\!\;\Gamma_{\mu\alpha\sigma}(-Q,Q,0)
  = \partial_{Q^\sigma} {\,}^*\!{\mit\Pi}_{\mu\alpha}(Q)
\end{equation}
where the vertex ${\,}^*\!\!\;\Gamma$ is the sum of the bare and
the hard-thermal-loop vertex and ${\,}^*\!{\mit\Pi}$ the sum of
the bare kinetic term and the hard-thermal-loop self-energy$\,$:
\begin{equation} \label{zusatz}
  {\,}^*\!{\mit\Pi}^{\mu\nu}  \; = \;
  Q^\mu Q^\nu - Q^2 g^{\mu\nu} + {\mit\Pi}^{\mu\nu}  \; = \;
  \left( {\mit\Pi}_t -Q^2 \right) A^{\mu\nu} + \left(
  {\mit\Pi}_\ell - Q^2 \right)  B^{\mu\nu} \;\; .
\end{equation}
Here $A$ is the projector on spatially transverse vectors
(s. (\ref{APi}) below) and
$B^{\mu\nu}=g^{\mu\nu} - Q^\mu Q^\nu / Q^2 - A^{\mu\nu}$.

The full next-to-leading order gluon self-energy is in general
gauge-parameter dependent and non-transverse, but one can show
on an algebraic level that the gauge parameter drops out from
$\delta {\mit\Pi}_{t,\ell}$ on the respective mass-shells and
also that $\delta {\mit\Pi}_{\mu\nu}$ is transverse (in the
four-dimensional sense) on the longitudinal mass-shell
\cite{flesh}. This on-shell gauge independence allows us to
further simplify the algebra by using Feynman gauge, where
indeed only the integral (\ref{S}) arises, and from on-shell
transversality we have
\begin{eqnarray} \label{APi}
   \delta{\mit\Pi}_t(Q) &=& {1\over 2} A^{\mu\nu}
   \delta {\mit\Pi}_{\mu\nu} = - {1\over 2} \left(
   \delta_{mn} - {q_m q_n \over q^2} \right)
   \delta {\mit\Pi}_{mn}  \\
     \label{BPi}
   \delta{\mit\Pi}_\ell(Q) &=& - {Q^2\over q^2}
     \delta{\mit\Pi}_{00} \;\; .
\end{eqnarray}

The whole algebra is now a matter of a few lines only:
\begin{eqnarray}
  \delta{\mit\Pi}_t(Q) &\simeq& - {g^2N\over 2} \left(
    \delta_{mn}  - {q_m q_n\over q^2}\right)
    {\,}^*\!\!\;\Gamma_{mai}(-Q,Q,0) \,
    {\,}^*\!\!\;\Gamma_{ani}(-Q,Q,0) \,
    {\cal S}_t(Q)  \nonumber \\
  &=& - g^2N \left( \partial_{q_i} \Delta^{-1}_t \right)^2 \,
      {\cal S}_t(Q) \nonumber \\
\label{t-S}
  &=&-g^2N\, 4q^2 \,\left[\, 1 + \partial_{q^2}
  {\mit\Pi}_t(\omega,q) \,\right]^2 \, {\cal S}_t(Q)
\end{eqnarray}
and
\begin{eqnarray}
   \delta{\mit\Pi}_\ell(Q) &\simeq& -g^2N\left( Q^2 \over
   q^2\right)^2 {\,}^*\!\!\;\Gamma_{00i}^2(-Q,Q,0) \,
   {\cal S}_\ell(Q)  \nonumber \\
   &=& - g^2N \left( \partial_{q_i} \Delta^{-1}_\ell \right)^2
       {\cal S}_\ell(Q) \nonumber \\
 \label{ell-S}
   &=& -g^2N\, 4q^2 \,\left[\, 1 + \partial_{q^2} {\mit\Pi}_\ell
   (\omega,q) \,\right]^2 {\cal S}_\ell(Q) \;\; .
\end{eqnarray}
This exhibits the prominent role played by the factors
${\cal S}_i$, $i=t,\ell$, studied in the previous section.
As they have no singular real parts at positive $q^2$ and
no singular imaginary part at $q^2<0$, (\ref{t-S}) and
(\ref{ell-S}) tell us that there is {\em no} danger from
the infrared in the corresponding parts of the polarization
functions at next-to-leading order. For the singular parts, on
the other hand, ${\cal S}$ contains in one simple expression
the essentials on both next-to-leading-order damping and
screening for $q^2>0$ and $q^2<0$, respectively (see Figure 1).

Let us first consider the consequences for propagating
quasi-particle excitations which have $\omega >m$ and $q \sim
m \gg \mu$. Inserting the on-shell values of the derivatives
of the various ${\mit\Pi}_i$ we obtain
\begin{eqnarray} \label{dpt}
 \delta{\mit\Pi}_t &\simeq&-ig^2NT{q\over 2\pi} \,\left[\, 1 +
 \partial_{q^2} {\mit\Pi}_t(\omega,q) \,\right]\,
 \ln\left( {1\over g} \right) \nonumber \\
 &=& -ig^2NT {3\over 4\pi} \left( {m^2\omega^2 \over Q^2}
     - Q^2\right) {1\over q} \ln\left({1\over g}\right) \\
 \label{dpl}
    \delta{\mit\Pi}_\ell(Q) &\simeq& -ig^2NT{q\over 2\pi}
    \,\left[\, 1+\partial_{q^2} {\mit\Pi}_\ell(\omega,q)
    \,\right]\, \ln\left( {1\over g}\right) \nonumber \\
 &=& -ig^2NT{3\over 4\pi} \left( m^2-Q^2 \right) {1\over q}
 \ln\left( {1\over g}\right) \;\; ,
\end{eqnarray}
where in evaluating $\cal S$ we have assumed $m/\mu\sim 1/g$
and used that $\sigma'=+1$ for both modes.

In order to derive corrections to the dispersion laws, one has
to take into account that the leading-order ${\mit\Pi}_{t,\ell}$
also varies when $\omega_{t,\ell}(q)$ changes. Expanding the
condition $\omega^2_{t,\ell}(q) = q^2 + {\mit\Pi}_{t,\ell} +
\delta{\mit\Pi}_{t,\ell}$ around the leading-order result
$\omega_0(q)^2$, we have
\begin{eqnarray} \label{o2tell}
  \delta\omega_{t,\ell}^2 \equiv
    (\omega^2 - \omega_{0}^2)_{t,\ell}
  &=& \delta{\mit\Pi}_{t,\ell} (\omega_0,q)
    + \delta\omega_{t,\ell}^2 \partial_{\omega_0^2}
    {\mit\Pi}_{t,\ell}(\omega_0,q) + O\left( \delta^2\right)
      \nonumber \\
  &=& {\delta{\mit\Pi}_{t,\ell} (\omega_0,q)
  \over 1 - \partial_{\omega_0^2}
  {\mit\Pi}_{t,\ell}(\omega_0,q)}+O\left( \delta^2\right) \;\; .
\end{eqnarray}

Interpreting the imaginary correction $\delta\omega^2
= -2i\omega_0 \gamma$ as a damping term $\gamma$, we obtain
(dropping the index 0 on $\omega$ in the final results, since
the difference is of higher order now)
\begin{eqnarray} \label{gtell}
  \gamma_{t,\ell} &=& {g^2NT\over 4\pi}\; {q \,\left[\, 1 +
   \partial_{q^2} {\mit\Pi}_{t,\ell}(\omega,q) \,\right]\,
   \over \omega \,\left[\, 1-\partial_{\omega^2}
   {\mit\Pi}_{t,\ell}(\omega,q)  \,\right]\, } \,\ln\left(
   {1\over g}\right) \nonumber \\
   &=& {g^2NT\over 4\pi}\, v_{t,\ell} \,\ln\left( {1\over g}
       \right)
\end{eqnarray}
with
\begin{equation} \label{vs}
v_\ell    = {\omega\over q}\; {3(m^2-Q^2)\over 3m^2-Q^2}
      \qquad \mbox{and} \qquad
v_t = {\omega\over q}\; {3(m^2\omega^2 - Q^4) \over
       3m^2 \omega^2-Q^4} \;\; ,
\end{equation}
both of which are zero at $q=0$ and approach 1 with increasing
$q$. This exactly reproduces and confirms the results of
Ref.~\cite{Pmov}.

The above calculations allows us equally to derive the
correction to the part of the dispersion laws which describe
screening of external perturbations with frequencies below the
plasma frequency, $\omega <m$ (see Figure 1). There the gluon
propagator exhibits poles for imaginary values of $q$, which
gives the effective screening length for a given frequency
$\omega$. In the static limit this reduces to the electric
(Debye) mass and the magnetic mass (which is zero at leading
order) for mode $\ell$ and $t$, respectively. In the general
case, both screening lengths are finite (which is sometimes
referred to as ``dynamical screening''). Here it is more natural
to keep $\omega$ fixed and to determine $\delta q(\omega)$
according to
\begin{equation} \label{q2tell}
  \delta(-q^2)_{t,\ell} \;=\; {\delta {\mit\Pi}_{t,\ell}
  (\omega,q)  \over 1 + \partial_{q^2}
  {\mit\Pi}_{t,\ell}(\omega,q)}+O\left( \delta^2\right) \;\; ,
\end{equation}
with $q(\omega)$ as given by the leading-order results.

Together with the results for ${\cal S}_{t,\ell}$ for imaginary
$q$, this gives for $|q|\sim m\gg \mu$
\begin{equation} \label{qscr}
 \delta|q|_{t,\ell} = {g^2NT\over 4\pi} \ln\left(
 {1\over g}\right) \;\; ,
\end{equation}
so the correction is now a simple constant. In the static limit
$|q|_\ell = \sqrt{{3} \,}^{\hbox to0.2pt{\hss$\vrule
height 2pt width 0.6pt depth 0pt $} \;\!} m = m_{\rm el.}$
and (\ref{qscr}) agrees with the next-to-leading order result
for the electric (Debye) screening mass of
Refs.~\cite{AKR,AKR2,BrN}.

In the previous section we have seen that the logarithmic
singularity of
$$ g^2{\cal S}\sim (g^2T/q) \ln \left( q/\mu \right)
   \sim g\ln\left( 1/g\right) \qquad\hbox{for $|q|\sim m$} $$
turns into a linear one for $|q|\to0$:
$$ g^2{\cal S}\sim gm/\mu \qquad\hbox{for \ $|q|
    \mbox{\,\raisebox{.3ex}{$\,<$}$\!
    \!\!\!\!\!$\raisebox{-.9ex}{$\,\sim\;$}} \mu$} \;\; , $$
which for $\mu\sim g^2T$ is $O(1)$. The final results for the
gluon self-energy (\ref{dpt},\ref{dpl}) come with a prefactor
that vanishes like $q^2$, so that both, the damping coefficient
and the correction to dynamical screening, as given above, do
not receive contributions from the singular integral ${\cal S}$.
This could change, however, when higher-order corrections are
taken into account, and we shall resume this point after
extending the above calculations to the fermionic spectrum.

%
%

\vspace{1.5cm} \hspace{.08cm} \parbox{15cm}{{\bf
4. \ Quark self-energy }
\vspace{1.1cm} } \hfill \vphantom{a} \nopagebreak \indent

The calculation of the infrared-singular parts of the one-loop
resummed quark self-energy can be simplified in full analogy
to the gluonic case of the previous section. Again the result
factorises into an expression involving only the external
momentum and the singular integral ${\cal S}_f$ introduced
in sect.~2. The dressed quark-quark-gluon vertex is needed
with vanishing gluon momentum only and can again be derived
from the self-energy through a differential Ward identity,
\begin{equation} \label{fermv}
  {\,}^*\!\!\;\Gamma_\mu(-Q,Q,0) = \partial_{Q^\mu}
  \left( Q\!\llap{/} - \Sigma(Q)\right)\;\; ,
\end{equation}
so that
\begin{equation}
 \delta \Sigma \,=\, - g^2C_F
  \,\left[\, \partial_{q_m} \left( Q\!\llap{/}
    - \Sigma(Q) \right) \,\right]\, \left( Q\!\llap{/}
    - \Sigma(Q) \right)  \,\left[\, \partial_{q_m} \left(
 Q\!\llap{/} - \Sigma (Q) \right) \,\right]\, {\cal S}_f \;\; .
\end{equation}

The corrections to the two branches of the fermionic dispersion
laws (see Figure 1) are determined by
\begin{equation}
  \delta {\mit\Pi}_\pm={1\over 4}{\rm tr}\left( (\gamma^0\mp
  \hbox{$q$\llap{/}} /q)\delta\Sigma\right) \;\; .
\end{equation}
Introducing
\begin{equation}
   D_\pm = \omega \mp q - {\mit\Pi}_\pm
\end{equation}
we obtain
\begin{equation} \label{dpialg}
  \delta{\mit\Pi}_\pm = -\left \{ D_\mp (\partial_q D_\pm)^2 +
  {1\over 2} D_\pm (D_+-D_-)^2 \right\} {\cal S}_f \;\; ,
\end{equation}
where only the first term in the curly brackets
survives on the respective mass-shell $D_\pm=0$.

In the fermionic case, the propagator has poles only for real
values of $q$, corresponding to the fact that fermionic quantum
numbers cannot get screened. For $q\sim M \gg \mu\sim g^2T$ we
have
\begin{eqnarray} \label{dpf}
  \delta{\mit\Pi}_\pm &=& -i\, {g^2C_FT\over 4\pi} \,
     |\partial_q D_\pm| \,\ln\left( {1\over g}\right)
     \nonumber \\
  &=& -i\, {g^2C_FT\over 4\pi} \,\left| 1\mp{2\omega
     (Q^2-M^2) \over qQ^2} \right|  \,\ln\left( {1\over g}
     \right) \;\; ,
\end{eqnarray}
where in evaluating ${\cal S}_f$ we have used
\begin{equation}
 1 + \partial_{q^2} {\mit\Pi}_f \;=\;
    - \,{1\over 2q} \,\partial_q \left( D_+ D_- \right)
\end{equation}
and ${\rm sign}(v_\pm) = -{\rm sign} ( \partial_q D_\pm)$
in evaluating ${\cal S}_f$. Determining the correction term to
the dispersion law $\omega_\pm(q) = \omega_{0\pm}(q) +
\delta\omega_\pm(q)$ requires again to take into account the
induced variation of the leading-order self-energy through
$\delta\omega_\pm$ according to
\begin{equation} \label{opm}
  \delta \omega_\pm = {\delta{\mit\Pi}_\pm(\omega_0,q) \over
     1 - \partial_{\omega_0}{\mit\Pi}_\pm(\omega_0,q)}
   = {\delta{\mit\Pi}_\pm(\omega_0,q) \over
     \partial_{\omega_0}D_\pm(\omega_0,q)} \;\; .
\end{equation}
Since $\delta{\mit\Pi}\pm$ is purely imaginary, this yields
a fermionic damping coefficient
\begin{eqnarray} \label{gpm}
  \gamma_\pm&=&{g^2C_FT\over 4\pi}\; {|\partial_q D_\pm| \over
      \partial_{\omega} D_\pm(\omega,q)} \;\ln\left({1\over g}
      \right)  \nonumber \\
  &=& {g^2C_FT\over 4\pi}\, |v_\pm|\, \ln\left( {1\over g}
      \right)  \nonumber \\
  &=& {g^2C_FT\over 4\pi}\, \left| {\omega(Q^2-M^2)\mp q
      Q^2 \over qM^2} \right| \,\ln\left( {1\over g} \right)
\end{eqnarray}
in accordance with the results of Ref.~\cite{Pmov}.

The group velocity $v_\pm$ equals $\pm1/3$ in the limit $q\to0$,
monotoneously increasing towards $+1$ for large $q$ (with a zero
for the plasmino mode at $q\approx 0.41M$). For $q\to0$,
(\ref{gpm}) is no longer valid but becomes proportional to $q/\mu$
in place of $\ln(1/g)$, so that the infrared-singular contribution
disappears. This agrees with the calculations performed in
Ref.\ \cite{KKM} at strictly $q=0$, where a finite result without
$\ln(1/g)$ was obtained.

%
%

\vspace{1.5cm} \hspace{.08cm} \parbox{15cm}{{\bf
5. \  Higher-order contributions }
\vspace{1.1cm} } \hfill \vphantom{a} \nopagebreak \indent

For $q\not=0$, we had to invoke higher-order corrections for
providing a physical infrared cutoff for the mass-shell
singularities in the next-to-leading order corrections to the
various on-shell self-energy components. Since the singularities
were only logarithmic, the assumption that such an infrared
cutoff will indeed be put into effect by the higher-order
contributions at a certain scale $\mu$ suffices to fix the
dominant term $\propto g\ln(1/g)$. Clearly, the other terms of
$O(g)$ depend on the details of the new infrared physics at the
scale $\mu$ and thus are beyond the perturbative scheme of
Braaten and Pisarski, which is based on a resummation of the
leading-order terms pertaining to the scale $m\sim gT$.
Only the real-part corrections to the dispersion laws at $O(g)$
are infrared-safe and calculable this way \cite{HS,flesh}.

In the long-wavelength-limit, $q\to0$, this sensitivity to
scales $\ll m$ disappeared because the potentially singular
terms in the self-energy corrections were of the form
\begin{eqnarray} \label{dpitell}
  \delta{\mit\Pi}_{t,\ell}(\omega,q) &\simeq& q^2 \,f_{t,\ell}
    (\omega/q)\, g^2 \, {\cal S}_{t,\ell} \\ \label{dpipm}
  \delta{\mit\Pi}_\pm(\omega,q) &\simeq& q\, f_\pm(\omega/q)\,
  g^2\, {\cal S}_f \;\; .
\end{eqnarray}
In sect.~2 we have observed that $\cal S$ becomes {\em linearly}
singular for $q\to0$, and assuming that higher-order corrections
render $\cal S$ finite so that ${\cal S} \sim T/\mu$, one can
indeed drop the above contributions at $q=0$.

However, this may cease to be justified if there are
higher-order corrections to the prefactor of ${\cal S}$ that do
not vanish when $q\to0$. In the resummed perturbation theory,
one collects all the contributions at the soft scale $gT$,
consistently disregarding potential terms in the effective
action that are proportional to $\mu/m$ and therefore suppressed
by extra powers of $g$. So higher-order corrections to the
dressed vertices and propagators could in principle change the
prefactor in (\ref{dpitell}) like
\begin{equation}
  q^2 \to q^2 + c_1 gmq + c_2 gm\omega + c_3 g m^2 + \ldots
\end{equation}
and in (\ref{dpipm}) like
\begin{equation}
  q \to q + d_1 gm + \ldots \;\; ,
\end{equation}
with dimensionless functions $c_i(\omega,q)$ and
$d_i(\omega,q)\,$.

If $\mu$ which cuts off the linear singularity of ${\cal S}$
for $q \mbox{\,\raisebox{.3ex}{$\,<$}$\!
       \!\!\!\!\!$\raisebox{-.9ex}{$\,\sim\;$}} \mu$
was much smaller than $gm$, then there would even be the
possibility that the {\em leading-order} results might get
modified by the higher-loop orders, but only for
$q \mbox{\,\raisebox{.3ex}{$\,<$}$\!
   \!\!\!\!\!$\raisebox{-.9ex}{$\,\sim\;$}} \mu$.
We shall exclude this rather improbable eventuality by assuming
that $\mu\sim g m \sim g^2T$.

Let us first consider the effect of nonzero $c$'s on the result
for dynamical screening, eq.~(\ref{q2tell}), for
$q \mbox{\,\raisebox{.3ex}{$\,<$}$\!
   \!\!\!\!\!$\raisebox{-.9ex}{$\,\sim\;$}} \mu$.
The correction to the inverse screening length $|q|$ becomes
\begin{eqnarray} \label{qscrc}
  \delta|q|_{t,\ell} &\sim& {gmq\over \mu} + c_1
  {g^2m^2 \over \mu} + c_2 {g^2m^2\omega\over q\mu}
  + c_3 {g^2 m^3 \over q\mu} + \ldots
  \nonumber \\
  &\sim& gq + c_1 gm + c_2 {gm\omega \over q}
  + c_3 {gm^2 \over q}  + \ldots
\end{eqnarray}
Obviously,
$c_2(\omega,q)$ and $c_3(\omega,q)$ should vanish for $q\to0$
in order that the kinematical situation for $q\to0$ does not
become singular, which we shall take for granted.

In the static limit, the transverse branch of the dispersion
laws has $|q|=0$ at leading-order, i.e. a vanishing magnetic
mass. A nonvanishing $\delta|q|_t$ for $\omega\to0$ would be
interpreted as the generation of a magnetic mass $m_{\rm magn}$.
The result obtained within the Braaten-Pisarski scheme vanishes
in this limit, which is in agreement with the null result of
Ref.~\cite{AKR}. But a nonzero $c_1$ would render
$m_{\rm magn}\sim c_1 gm$, which is consistent with our
assumption that the cut-off $\mu$ is of the order of the
magnetic mass (although not necessarily identical with it).
The linear mass-shell singularity of $\cal S$ for
$q \mbox{\,\raisebox{.3ex}{$\,<$}$\!
   \!\!\!\!\!$\raisebox{-.9ex}{$\,\sim\;$}} \mu$ could thus
play a prominent role for the generation of a magnetic screening
mass through higher-order corrections.

Turning now to the propagating gluonic modes, we note that
for $q\to0$, the singular contributions to the damping constant
still vanish like
\begin{equation}
   \gamma_{t,\ell} \;\sim\; {gq^2\over \mu}
   + c_1 {g^2qm \over \mu} + \ldots \;\; ,
\end{equation}
since we have excluded $c_2$ and $c_3$ for $q\to0$. Thus there
seems to be little danger that the relative-order-$g$ results
that have been obtained previously for $q=0$ \cite{BPpd,HS}
could be modified by higher-order corrections.
Also for nonzero
$q \mbox{\,\raisebox{.3ex}{$\,<$}$\!
   \!\!\!\!\!$\raisebox{-.9ex}{$\,\sim\;$}} \mu$,
the contribution from the mass-shell singularities remains
below $O(gm)$.

The situation is somewhat different for the fermionic modes,
however. There we have
\begin{equation}
  \gamma_\pm \; \mbox{\,\raisebox{.3ex}{$\,<$}$\!
    \!\!\!\!\!$\raisebox{-.9ex}{$\,\sim\;$}}\; {g q m \over
    \mu} + d_1 {g^2 m^2 \over \mu} + \ldots \;\; ,
\end{equation}
and already the first term, which arises within the
Braaten-Pisarski scheme, is of the order $gm\sim g^2T$ for
$q\sim \mu$. For such momenta $\gamma$ is obviously not
calculable within the one-loop resummed approximation, because
through the linear divergence of $\cal S$, loop momenta of the
order of $\mu$ contribute on a par with the ones of order $m$.
For $q\to0$, these contributions are suppressed,
but if higher-order terms could produce a nonvanishing $d_1$,
then also the strict $q=0$ result for $\gamma$ at order $g^2T$
would become infested by higher-order contributions.

However, the proportionality of $\delta{\mit\Pi}_\pm$ to $q$ can
be traced back to the Ward identity (\ref{fermv}) which is
responsible for the simple form of (\ref{dpialg}): therein
both $D_+$ and $D_-$ vanish like $q$ for $q\to0$. In the Abelian
case, the tree-level-like Ward identities hold also beyond the
level of hard thermal loops, whereas in the nonabelian case, one
can retain simple Ward identites by choosing to work in axial
gauge, so it seems plausible that $d_1\propto q$, which makes
the results obtained for strictly $q=0$ in Ref.~\cite{KKM}
stable against higher-order contributions, despite the lurking
mass-shell singularities.

%
%

\vspace{1.5cm} \hspace{.08cm} \parbox{15cm}{{\bf
6. \ Conclusion }
\vspace{1.1cm} } \hfill \vphantom{a} \nopagebreak \indent

We have found that the next-to-leading order corrections
to dynamical screening masses for gluonic fields with frequencies
below the plasma frequency as well as the damping constants for
the propagating quasi-particle modes are strongly sensitive to
the magnetic-mass scale except when the wave-vector is exactly
zero. On the other hand, the next-to-leading order corrections
to the real part of the dispersion laws of the propagating modes
have turned out to be infrared-safe.

The infrared singular contributions for both, screening and
damping, are determined in essence by one simple expression
exhibiting (quasi-particle) mass-shell singularities. For
large enough modulus of the wave-vector the latter are
logarithmic, yielding contributions of relative order
$g \ln(1/g)$. The coefficient in front of the logarithm is
calculable once the scale of the cutoff brought about by
higher-order contributions has been determined. The
coefficients under the logarithm $\ln(1/g)$, which are of
obvious importance when the coupling constant is not
infinitesimally small, are clearly beyond these perturbative
considerations.

For small wave-vector, we have found that the mass-shell
singularities become even linear, which in principle opens a
way for higher-order corrections to contribute on a par with
the ones obtained within the resummed perturbation theory of
Braaten and Pisarski. We have argued that it is plausible that
the case of exactly vanishing wave-vector is stable against
such corrections, whereas the particular case of the damping
constant of fermionic modes with nonzero wave-vector of the
order of the magnetic mass-scale remains uncalculable.

Linear mass-shell singularities have previously been identified
as the root of a potential problem with gauge independence
\cite{BKS} of next-to-leading order corrections to the dispersion
laws. Whereas formally one can prove gauge fixing independence
\cite{KKR}, in covariant gauges the unphysical modes of the gluons
behave like zero-mass particles, and lead to linear divergences
in the residues of the quasi-particle propagators. Unless they are
identified as such by the introduction of a (purely technical)
infrared cut-off, they can mimic contributions to the pole
position \cite{BKSC}.

However, the type of mass-shell singularities that we have
discussed in this paper appear directly in the corrections to
the pole position. They call for a physical cut-off to be
provided by higher-order corrections. Indeed, in the results
for dynamical screening of transverse gluonic modes, we have
seen that the linear mass-shell singularities themselves could
play an important role in a dynamical generation of a magnetic
screening mass $\sim g^2T$, which would be the most obvious
candidate for such a physical infrared cut-off.

In the nonabelian case, the mass-shell singularities are there
even in the purely static situation, where they provide the
dominant next-to-leading order contribution to the Debye
screening mass. Here the other possibility for an effective
infrared cut-off that has been discussed in the literature,
namely damping of the internal propagators, can hardly be
operative, which underlines the need of (chromo-)magnetostatic
screening.

On the other hand, in the Abelian case, a magnetic screening
mass cannot be generated. This is no problem for the Debye mass,
which is infrared safe in QED. But mass-shell singularities are
there in the next-to-leading order corrections to the electron
propagator, and there has been some controversy
\cite{BNN,Smilga,PPS} on whether the finite width of the
internal propagators alone can provide the necessary cut-off
$\mu$, in particular when the singular contributions are
evaluated on the corrected quasi-particle mass-shells, i.e.
including damping. This has been cleared up recently in
Ref.~\cite{BK}. Our approach, however, was a strictly
perturbative one. Through eq.~(\ref{opm}) it requires to
evaluate the corrections at the location of the (real)
leading-order position, which, at this level, is in accordance
with the findings of Ref.~\cite{BK}. Our results and
conclusions should therefore be quite independent of the details
of the actual higher-order effects, as long as they are indeed
able to cut off the quasi-particle mass-shell singularities.

{\bf Acknowledgments:} F.\ F.\ is supported by Deutsche
Forschungsgemeinschaft (DFG).

%
%
              \let\dq=\theequation
\renewcommand{\theequation}{A.\dq} \setcounter{equation}{0}

\vspace{1.5cm} \hspace{.08cm} \parbox{15cm}{{\bf
Appendix }
\vspace{1.1cm} } \hfill \vphantom{a} \nopagebreak \indent

In the following we describe some of the steps which are
encountered when deriving the main results of sects. 2 and 3
in a more pedestrian way.

The complete next-to-leading order correction to the
longitudinal gluon polarization function $\delta {\mit\Pi}_\ell$,
on the longitudinal mass-shell, has recently been written down
in full detail in eq.~(4.5) of Ref.~\cite{flesh}, albeit still
at a purely algebraic level. To cover also the transverse case,
we generalize the result obtained in \cite{flesh} and give an
expression for $\delta{\mit\Pi}^{\mu\nu}$ which is valid when
used under the trace {\em either} with the matrix $B_{\mu\nu}$
introduced in sect.~3 and taken at the longitudinal mass-shell
{\em or} with $A_{\mu\nu}$ and taken at the transverse
mass-shell:
\begin{equation} \label{A1}
   \delta{\mit\Pi}^{\mu\nu} =  \, g^2 N \;
\sum \left(
    c_0^{\mu\nu}  +  \Delta_\ell^- \Delta_\ell^{} \,
    c_{\ell\ell}^{\mu\nu}
    + \Delta_\ell^- \Delta_t^{} \, c_{\ell t}^{\mu\nu}
    + \Delta_t^- \Delta_t^{} \, c_{tt}^{\mu\nu} \right) \qquad
\end{equation}
with the coefficient matrices $c^{\mu\nu}$
\begin{eqnarray} \label{A2}
 c_0^{\mu\nu} &=& \Delta_0^- \Delta_0 \,\left[\, 2 P^2
     g^{\mu\nu} - 4 P^\mu P^\nu \,\right]\,  \\
 \label{A3}
 c_{\ell\ell}^{\mu\nu} &=& {P^2 K^2 \over 2p^2 k^2 }
   {\,}^*\!\!\;\Gamma^{\mu 0 0} {\,}^*\!\!\;\Gamma^{\nu 0 0}
   - {P^2 \over 2 p^2}\, \delta_\ell^-
   \,{\,}^*\!\!\;\Gamma^{\mu \nu 0 0}  \\
 \label{A4}
 c_{\ell t}^{\mu\nu} &=& - \, {P^2 K^2 \over p^2 k^2 }
   {\,}^*\!\!\;\Gamma^{\mu 0 0} {\,}^*\!\!\;\Gamma^{\nu 0 0}
   - {K^2 \over k^2} {\,}^*\!\!\;\Gamma^{\mu 0 \rho}
     {\,}^*\!\!\;\Gamma^{\nu 0}_{\;\;\;\,\rho}
   + {P^2 \over 2 p^2}\, \delta_\ell^- \,
     {\,}^*\!\!\;\Gamma^{\mu \nu 0 0} \\
 \label{A5}
 c_{tt}^{\mu\nu} &=& {P^2 K^2 \over 2p^2 k^2 }
   {\,}^*\!\!\;\Gamma^{\mu 0 0} {\,}^*\!\!\;\Gamma^{\nu 0 0}
    + {K^2 \over k^2} {\,}^*\!\!\;\Gamma^{\mu 0 \rho}
      {\,}^*\!\!\;\Gamma^{\nu 0}_{\;\;\;\,\rho}
    + {1\over 2} {\,}^*\!\!\;\Gamma^{\mu \rho \lambda}
      {\,}^*\!\!\;\Gamma^{\nu}_{\;\;\rho \lambda} \nonumber \\
    & & {}    - \, 3 g^{\mu\nu} \delta_t
    + \delta_\ell^{-} \delta_\ell^{}\, {P_0 K_0 \over P^2 K^2 }
      U^\mu U^\nu + {1 \over 2 p^2 k^2 } \, \left[\,  R^\mu_{}
      R^\nu_{-} + R^\mu_{-} R^\nu_{} \,\right]\, \;\; . \quad
\end{eqnarray}
The notation is that of \cite{flesh}$\,$:
$K = Q-P$, ${\bf k} = {\bf q} - {\bf p}$,
$\Delta_0 = 1/P^2$, $\Delta_i = 1 / (P^2- {\mit\Pi}_i (P) )$
$(i=\ell,\, t)$,
$\delta_i = P^2 - {\mit\Pi}_i (P) = \Delta_i^{-1}$,
$U = (1 , {\bf 0} )\,$. An index minus refers to the shift
$P \rightarrow K$ in the corresponding quantity. The four
vector $R$ has no zeroth component$\,$:
$R^\mu = \big( \delta_t - \delta_\ell P_0^2 / P^2 \big)
 \big( P_0 U^\mu - P^\mu \big)$. The arguments of
the 4--leg vertex $\!{\,}^*\!\!\;\Gamma$ are $Q,-Q,-P,P\,$,
those of all 3--leg $\!{\,}^*\!\!\;\Gamma$'s are $Q,-K,-P\,$.
Finally, the unadorned sum symbol is short-hand for
$\sum_{P_0}\int{d^3p/(2\pi)^3}$.

The above result (\ref{A1}) has a convergent sum over
$P_0$, and its $p$-integration is restricted to soft
values automatically due to the subtraction of the
hard contribution. The calculation was done in general
covariant gauge with gauge parameter $\alpha$, which
dropped out algebraically.

As in sect.~3, we concentrate on contributions involving a
transverse propagator,
\begin{equation} \label{Aterm}
  \sum \Delta_t \,\Delta_i^{-} \; c(P_0 , {\bf p})
\end{equation}
Using the spectral representation
\begin{equation} \label{Adens}
  \Delta_t = \int_{-\infty}^\infty \! dx \, x \,
  {\,\rho_t (x,p) \over P_0^2 - x^2 } \;\; ,
\end{equation}
we may write
\begin{equation} \label{Aspect}
  \sum \Delta_t \,\Delta_i^{-} \, c =
  \int\! {d^3 p \over (2\pi)^3 } \int_{-\infty}^\infty \! dx
  \, {1 \over x}\, \rho_t (x,p) \sum_{P_0}
  {x^2 \over P_0^2 - x^2 } \,
  \Delta_i \left( Q_0 - P_0 , {\bf q} - {\bf p} \right)
  \, c(P_0, {\bf p} ) \;\; .
\end{equation}
Inspecting now the integration region of small $p$, one
realizes that the weight of the transversal (but not the
longitudinal) density (times $1/x$) is concentrated at $x=0\;$
(see e.g. eq.~(B.13) of Ref.~\cite{HS}\,)\,:
\begin{equation} \label{Adelta}
    {1\over x} \,\rho_t (x,p) \;\rightarrow\; {1\over p^2}
    \,\delta (x) \qquad \quad (\, p^2 \ll m^2 \, ) \;\; .
\end{equation}
Inserting (\ref{Adelta}) in (\ref{Aspect}) and performing the
$x$-integration, one finds that the sum over $P_0$ reduces
to the term $P_0=0$ provided $c(P_0, {\bf p} )$ has no poles
at $P_0=0$, but this is easily excluded by inspection.
Infrared singularities can now occur when $\Delta_i^{-}$
or $c(0,{\bf p})$ diverges for $p\to0$. By inspection one
finds that $c(0,{\bf p})$ is regular, so all singularities
are due to the mass-shell singularities when approaching
mass-shell $i$. The first term of the Taylor series for
$c(0,{\bf p})$ is responsible for the dominant contribution,
which thus involves only the paradigmatic term ${\cal S}_i$
studied in sect.~2 times functions of the external momentum.
This way eqs.~(\ref{t-S}) and (\ref{ell-S}) are readily
reproduced.

%
%
                    \renewcommand{\section}{\paragraph}

\end{document}